# A REDUCED OFFSET BASED METHOD FOR FAST COMPUTATION OF THE PRIME IMPLICANTS COVERING A GIVEN CUBE

Fatih BASCIFTCI [1], Sirzat KAHRAMANLI [2]

[1] Department of Electronics and Computer Education, [2] Department of Computer Engineering, Selcuk University, Konya, Turkey

{basciftci, sirzat }@selcuk.edu.tr

**Abstract.** In order to generate prime implicants for a given cube (minterm), most of minimization methods increase the dimension of this cube by removing one literal from it at a time. But there are two problems of exponential complexity. One of them is the selection of the order in which the literals are to be removed from the implicant at hand. The latter is the mechanism that checks whether a tentative literal removal is acceptable. The reduced Offset concept has been developed to avoid of these problems. This concept is based on positional-cube representation where each cube is represented by two n-bit strings. We show that each reduced Off-cube may be represented by a single n-bit string and propose a set of bitwise operations to be performed on such strings. The experiments on single-output benchmarks show that this approach can significantly speed up the minimization process, improve the quality of its results and reduce the amount of memory required for this aim.

# REDUCED OFFSET BASED METHOD FOR FAST COMPUTATION OF THE PRIME IMPLICANTS COVERING A GIVEN CUBE

Fatih BASCIFTCI [1], Sirzat KAHRAMANLI [2]

[1] Department of Electronics and Computer Education, [2] Department of Computer Engineering, Selcuk University, Konya, Turkey

{basciftci, sirzat }@selcuk.edu.tr

## 1. INTRODUCTION

Sum-of-products (SOP) minimization is a basic problem in logic synthesis [3,25]. It is also used for optimizing the care-networks when a design is carried out hierarchically [2,20] and for optimization of test generators [15,20]. SOP minimization is also very important for obtaining prime cubes containing source and target nodes and fixing shortest paths between them in hypercube configured systems [12,14,22]. However, due to the exponential nature of the exact SOP minimization problem, the state-of-the-art algorithms can typically handle functions with up to hundred product terms (cubes) in the minimum SOP [3]. Therefore, most of the practical applications and computer aided design (CAD) tools rely on direct-cover heuristic minimization methods [3].

Generally, the direct-cover heuristic minimization methods use the implicant expansion (reduction) concept to generate the set of prime implicants (PIs) covering the given cube (minterm) P. The function $f$ to be minimized by such a method is represented by the *Onset, Offset* and *Don't care set* that are the sets of minterms (cubes) making the function *f* equal to 1, equal to 0 and unspecified, respectively. We denote these sets by $S_{ON}$, $S_{OFF}$ and $S_{DC}$, and their cardinalities by $w(S_{ON})$, $w(S_{OFF})$ and $w(S_{DC})$, respectively. Similarly, the cardinality of any set $X$, introduced in sequel, we shall denoted by $w(X)$.

In order to obtain a minimum SOP for the given function, the typical direct-cover heuristic minimization method is realized by repeating the following steps until $S_{ON}$ is covered completely [3,8,16,21].

***Direct_Cover*** *// Input: $S_{ON}$, $S_{OFF}$; Output: A minimum form of the given function.*

1) An On-cube to be covered is chosen,
2) The set of PIs covering given minterm is generated,
3) The essential prime implicant (EPI) is identified,
4) A covering operation is performed.

In this algorithm the most time-consuming step is the second one. In [4,5] it is stated that this step is of polynomial complexity in number of variables (n). But our estimations show that this complexity is of degree that higher than polynomial.

Recall that an implicant is a product term that covers at least one minterm from $S_{ON}$ and does not cover any cube from $S_{OFF}$. Therefore, each implicant, which is expanded by removing any literal (variable or its complement) from it, must be intersected with the set $S_{OFF}$ to determine whether a tentative literal removal is acceptable. This process is known to be of polynomial complexity [4,5,24]. However, since $w(S_{OFF})=k \times 2^n$, where $k<1$, the total complexity of each PI construction process may be specified as a product of polynomial $(O(n^2))$ and exponential $(O(2^n))$ ones.

Note that the computational efficiency of the *Expand* procedure and the quality of the result (the cardinality of the final cover), generated by it, depend on two factors [8]:

1) The order in which the implicants are expanded
2) The order in which the literals are removed from the implicant

The rationale strategy for the first factor is to expand firstly those implicants that unlikely to be covered by other ones [8,9,10,13,16,17]. There are also several strategies for the second factor such as Sequential Search, Multiple Sequential Search, Distributed Multiple Sequential Search, and Distributed Exhaustive Implicant Expansion among which the first strategy is preferred [15]. Note that even some differences in implementation of the *Expand* operator would lead to different covers with different cardinalities [8,15,16,25]. Therefore to improve the quality of solutions, the programs such as *MINI, PRESTO* and *Espresso* iteratively manipulate the cover computed by operators *Expand*, *Reduce* and *Reshape* or *Irredundant* [8]. The algorithm terminates when the iteration of these operators does not reduce further the cover cardinality. Note that, none of the algorithms based on similar approaches is consistently better than the others for all logic functions. There are classes of functions where one heuristic algorithm is better than the others [8, 16].

In order to avoid problems, specific to the implicant expansion concept, *Abdul A. Malik, Robert K. Brayton, A. Richard Newton and Alberto Sagniovanni- Vincentelli* have developed the reduced Offset based minimization concept [1,2]. As will be shown below, an algorithm realizing this concept should be consisting of three steps, one of which is intractable due to its exponentional complexity. But on the other hand, due to own pure logical nature, the realization of the reduced Offset based minimization concept seems to be significantly speeded up by using a few transformations, specially developed for it.

The study is organized as follows. In section 2, the complexity of the reduced offset concept based algorithm is estimated. In section 3, the method of representation of the reduced off cubes by n-bit vectors and the method of generation of PIs by using these vectors are explained. In section 4, the results of experiments performed on 45 standard single-output MCNC benchmarks are showed. In section 5, the conclusion is given.

## 2. THE ESTIMATION OF THE COMPLEXITY OF THE REDUCED OFFSET CONCEPT BASED ALGORITHM

Recall that the reduced Offset concept has been developed to speed up the second step of the algorithm *Direct_Cover* explained in Section 1. According to this concept, the function to be minimized is represented by $S_{ON}$ and $S_{OFF}$. By special handling the elements of $S_{OFF}$ on the chosen On-cube *P* the reduced Offset $S_R(p)$ (that is valid only for *P*) is generated. The set $S_R(p)$ is minimized and the minimal set $S_{RM}(p)$ of the reduced cubes (RCs) is obtained. Then, by using DeMorgan's and Nelson laws the set $S_{RM}(p)$ is transformed into the set $S_{PI}(p)$ that contains all PIs covering the On-cube *P*. This method can be realized by the following three procedures [2].

*1. The Procedure Reduce_$S_{OFF}$*

This procedure transforms each cube $Z \in S_{OFF}$ into the corresponding RC $Z^r$ as follows [2,24].

$$If\ p_i = \bar{z}_i\ then\ c_i = z_i\ else\ c_i = x,\ \forall i \in \{0,1,\ldots,n-1\} \quad (1)$$

Where $P=p_{n-1}p_{n-2}\ldots p_0$ is the On-cube being processed, $Z=z_{n-1}z_{n-2}\ldots z_0$ is an Off-cube being reduced, $Z^r = c_{n-1}c_{n-2}\ldots c_0$ is the reduced form of the cube $Z$ and $x$ taking place in position of the variable $z_i$ means that this variable is *do not care* for $Z^r$. As it is easily seen, the procedure *Reduce* $S_{OFF}$ transforms the set $S_{OFF}$ into a set $S_R(p)$ by removing all literals from the cubes of $S_{OFF}$ except those that are complements of the appropriate literals in $P$ [2]. We have estimated the complexity of the procedure *Reduce_*$S_{OFF}$ via the complexity of a sub procedure that realizes the reduction of one Off-cube by 6 bitwise operations. Since this sub procedure must be repeated for all Off-cubes, the time complexity of the procedure *Reduce_*$S_{OFF}$ is to be $6 \times w(S_{OFF})$ computer's instruction cycles (CICs).

*2. The procedure Minimize_*$S_R(p)$

This procedure removes from $S_R(p)$ all cubes absorbed by other ones [2,23]. The work of this procedure may be formally expressed as follows.

$\forall (i \in \{1,2,\ldots, w(S_{OFF})-1\}, j \in \{2,\ldots, w(S_{OFF})\}),$

$$(Z_i^r, Z_j^r) = \begin{cases} Z_j^r, \text{ if } Z_i^r \cap Z_j^r = Z_i^r \\ Z_i^r, \text{ if } Z_i^r \cap Z_j^r = Z_j^r \\ (Z_i^r, Z_j^r), \text{ if } Z_i^r \cap Z_j^r \notin \{Z_i^r, Z_j^r\} \end{cases} \quad (2)$$

We are estimating the complexity of this procedure by using complexity of the sub procedure *Detect_Absorbed* which realizes the rule 2 for a certain pair (i, j). This sub procedure contains 3 bitwise, 2 conditional and 3 return instructions. Since it must be applied $w(S_{OFF}) - j$ times for each $Z_j^r \in S_R(p)$, the procedure *Minimize_$S_R(p)$* is to be of polynomial (quadratic) time complexity.

*3. The procedure Generate_$S_{PI}(p)$*

The function realized by this procedure may be expressed as follows [7,12,18].

$$S_{PI}(p) = \{x\}^n \ \# \ \overline{S}_{RM}(p) = \overline{\overline{S}}_{RM}(p) = \overline{Z}_1^r \,\&\, \overline{Z}_2^r \ \&\ldots\&\, \overline{Z}_m^r, \qquad (3)$$

where $\{x\}^n$ is the n-dimensional universal cube. The formula (3) is realized as follows:

1) The set $S_{RM}(p)$ is transformed into a product-of-sums (POS) form by using DeMorgan's law,
2) The obtained POS is transformed into a SOP by using Nelson theorem.

***Example.*** Let $P=001$ and $S_{OFF} = \{000, 100, 111\}$. Find $S_{PI}(101)$.

There the transformations on the formulas (1), (2), and (3) are denoted by $T_1$, $T_2$ and $T_3$, respectively, in the solution of this example.

$T_1$: $(P, S_{OFF}) \rightarrow S_R(001) = \{xx0, 1x0, 11x\}$;

$T_2$: $S_R(001) \rightarrow S_{RM}(001) = \{xx0, 11x\}$;

T3: $S_{RM}(001) \rightarrow S_{PI}(101) = \overline{S}_{RM}(101) = \{\overline{xx0}\}$ & $\{\overline{11x}\} = \{xx1\}\&\{0xx, x0x\} = \{0x1, x01\}$

Our experiments over a lot of functions have shown that the cardinality of $S_{RM}(p)$ does not exceed 2.5n. Namely, the maximum number of clauses (maxterms) forming a POS is limited above by 2.5n. Since generally each clause contains at least two literals, the cardinality of SOP may reach of the number $O(2^{2.5n})$. In our opinion, this is one of the main reasons making the reduced Offset concept time-consuming when $S_{OFF}$ is unreasonable large and there are many On-cubes to be handled [2].

## 3. THE DIFFERENCE INDICATORS BASED REPRESENTATION OF THE REDUCED OFFSET AND GENERATION OF PRIME IMPLICANTS

### *3.1. Representation of the Reduced Cube by its Difference Indicator*

The reduced Offset concept, as most of minimization methods, uses positional-cube notation for representation of the cubes in computers. According to this notation, an uncomplemented variable $y_i$, a complemented variable $\bar{y}_i$ and a '*don't care variable*' (missing variable in a product term) are represented by bit–pairs 01, 10 and 11, respectively. The bit-pair 00 represents no value of the variable and therefore its presence in any variable position means that the cube at hand is empty and should be deleted [8,21]. If to denote the left and right bits of each bit-pair by *L* and *R*, respectively, then a cube $Z=z_n z_{n-1} \ldots z_1$: $\forall_i$, $z_i \in \{0,1,x\}$, may be represented by the pair $Z=(Z_L, Z_R)=(L_n, L_{n-1}, \ldots L_1, R_n, R_{n-1}, \ldots R_1)$ that is the most suitable representation of cubes for computers [11]. For example, the cube $Z=0x1x0$ will be represented as *$(Z_L,Z_R)=(11011, 01110)$*. Namely, in positional-cube notation, each cube is represented by two n-bit strings. But our studies have shown that each RC may be represented by a single n-bit string that allows us to significantly speed up the PIs generation process and reduce the amount of memory needed for this aim. Our starting point is as follows.

As it is well known, $w(S_{OFF})=w(S_R(p))=O(2^n)$, $\forall P \in S_{ON}$. But there most of the RCs in $(S_R(p))$ are absorbed by other fewer ones [8,23]. Our experiments performed on a lot of functions have shown that $w(S_{RM}(p)) \leq 2.5n$, $\forall P \in S_{ON}$. Namely, there at least $O(2^n)-2.5n$ RCs in $S_R(p)$ that should be removed from them. Hence, we try to represent the RCs in such a form that to be allowing us to speed up the operation obtaining which of two RCs being compared is redundant and to reduce the number of repetitions of this operation needed for generation of $S_{RM}(p)$. For this aim we use the following relation. The cube $Z_i$ is absorbed by the cube $Z_j$ if:

1) The set of don't care literals in $Z_i$ is a subset of ones in $Z_j$,

2) The same appearing literals in $Z_i$ and $Z_j$ have the same values.

Notice that, due to the formula (1), the second condition above is always satisfied for all pairs of RCs, and therefore it does not need to be checked. Thus we may use only one n-bit string per RC instead of two ones. Such a bit-string (BS) contains 0s and 1s in positions corresponding to the appearing and don't care literals of the RC represented by this BS, respectively. We call such a BS a literally *Difference Indicator* (DI). According to this approach, the DI for a cube $Z_j \in S_{OFF}$ may be generated by the following extremely simple procedure.

***Generate_ $D_j$*** *$(P, Z_{jR})$*

*Return $(D_j=P\oplus Z_{jR})$*

Where, $P$ is an On-cube (minterm) on which $D_j$ is generated, and $Z_{jR}$ is the string of the right bits of the cube $Z_j$. By processing all of cubes of $S_{OFF}$ by the procedure *Generate_$D_j$* the set of all DIs, denoted by $S_D(P)$, is generated. We prove the correctness of this approach via the procedure *Derive_ $Z_j^r$ $(P,D_j)$* that unambiguously transforms any $D_j$ into the appropriate RC $Z_j^r \in S_R(p)$.

***Derive_*** *$Z_j^r$ $(P,D_j)$*

*Return $(Z_{jL}^r = P \mid \sim D_j;\ Z_{jR}^r = \sim P \mid \sim D_j)$*

Where, ¦ is the bitwise OR operation symbol, and $Z_{jL}^r$ and $Z_{jR}^r$ are the strings of the left and right bits of the reduced cube $Z_j^r$, respectively.

***Example.*** $S_{ON}$={011, 101, 110 }, $S_{OFF}$={001, 110}. By using the DI approach, find the set $S_R$(p) for the minterm P=$P_R$=101.

1. *Application of the procedure Generate_$D_j$*

   *1.1. $D_1=P_R\oplus Z_{R1}=101\oplus 001=100$;*

   *1.2. $D_2=P_R\oplus Z_{R2}=101\oplus 110=011$*

   *Thus $S_D(101)=\{D_1,D_2\}=\{100, 011\}$*

2. *Application of the procedure Derive_$Z_j^r$*

*2.1.1.* $Z^{r}_{1L}$ *=P+~D₁=101+011=111*

*2.1.2.* $Z^{r}_{1R}$ *=~P+~D₁=010+011=011*

*2.1.3.* $Z^{r}_{1}=(Z^{r}_{1L}, Z^{r}_{1R})=(111,011)\rightarrow 0xx$

*2.2.1.* $Z^{r}_{2L}=P+D_2=101+110=111$

*2.2.2.* $Z^{r}_{2R}=\sim P+D_2=010+110=110$

*2.2.3.* $Z^{r}_{2}=(Z^{r}_{2L}, Z^{r}_{2R})=(111,110)\rightarrow xx0$

*Thus* $S_R(p)=\{Z^{r}_{1}, Z^{r}_{2}\}=\{0xx, xx0\}$

Notice that the procedure *Derive_*$Z^{r}_{j}$ is given here only for demonstration of the correctness of the DI approach. It will not be used for any other purpose.

### *3.2. The Formation of the Minimal Set of Difference Indicators*

Recall that according to reduced Offset concept, the set of reduced cubes $S_R(p)$ is generated, and then the minimal set of reduced cubes $S_{RM}(p)$ is formed by removing from $S_R(p)$ redundant (absorbed) cubes. As it is easy to see, this process is of polynomial complexity. Since in the DIs approach, instead of the sets $S_R(p)$ and $S_{RM}(p)$ the sets $S_D(p)$ and $S_{DM}(p)$ (the minimized form of $S_D(p)$) are generated, respectively. The transformation of $S_D(p)$ into $S_{DM}(p)$ is also to be of polynomial complexity. In order to avoid of the exponential cardinality set $S_D(p)$ and polynomial complexity problem of transformation of $S_D(p)$ into $S_{DM}(p)$, we try to form the set $S_{DM}(p)$ directly without any using the set $S_D(p)$. For this aim we have developed the procedure *Reform_*$S_{DM}(p)$ that compares each new generated $D_j$ to those that already in $S_{DM}(p)$. This comparison is continued until $D_j$ is absorbed or until all elements of $S_{DM}(p)$ are handled. Note that to generate the set $S_{DM}(p)$ completely, the procedures *Generate_*$D_j$ and *Reform_*$S_{DM}(p)$ must be applied to all elements of $S_{OFF}$ [2,23,24]. Namely, it must be repeated exactly $w(S_{OFF})$ times for each On-cube that needs to be processed. This may be done by the procedure *Generate_*$S_{DM}(p)$ given below.

***Generate_*$S_{DM}(p)$** $(P_R, S_{OFF}, w(S_{OFF}))$

$S_{DM}(p)=\{1\}^n, w(S_{DM})=1$ *// Set initial content and cardinality of* $S_{DM}(p)$

*For j=1 to m Do*

{ $D_j=P_R \oplus Z_j$ *// The body of the procedure Generate_*$D_j$

*Reform_*$S_{DM}(p)$ $(D_j, S_{DM}(p), w(S_{DM}))$

}

*Return* $(S_{DM}(p), w(S_{DM}))$

### *3.3. The Generation of the Bit-Vectors Representing the DeMorgan's Clauses*

According to the DI approach, the set of all prime implicants $S_{PI}(p)$ for an On-cube $P$ is generated by processing the set of difference indicators $S_{DM}(p)$. Let us to make the following definitions.

*Definition 1.* A DI of weight of m (containing m 1s) is called an *m-DI*.

*Definition 2.* The projection of the difference indicator $D=d_{n-1}d_{n-2}\ldots d_i\ldots d_0$ on the coordinate $i$ is expressed as follows [6].

$$D[i]=\underbrace{00...0}_{n-i-1}\, d_i\, \underbrace{0...00}_{i}\ , \quad d_i \in \{0,1\} \tag{4}$$

In order to handle DIs according to the DeMorgan's transformation of a product term into its clause equivalent, let us enumerate the elements of $S_{DM}(p)$ from $1$ to $w(S_{DM})$. Then, based on the formula (4) we may express the processing of each $D_j \in S_{DM}(p),\ j \in \{1,2,\ldots,\ w(S_{DM})\}$ by the rule (5) that may be realized by the procedure *Generate_M<sub>j</sub>(p)* given below.

$$M_j(p)=\{D_j[i]:\ d_i=1\},\quad i=1,2,\ldots,n \qquad (5)$$

***Generate_M$_j$(p)*** *(j, $D_j$, n)*

$M_j(p)=\varnothing;\ \ w(M_j)=0;\ \ B_0=D_j$

*While* $B_0 \neq 0$ *Do*

$\{\quad B_1=D_j-1;$

$\quad B_0=B_1 \& D_j;$

$\quad B_2=B_0 \oplus D_j;$

$\quad M_j(p)=M_j(p) \cup B_2;$

$\quad w(M_j)=w(M_j)+1$

$\}$

*Return* $(M_j(p),\ w(M_j))$

This procedure transforms the given *m-DI* (denoted by $D_j$) into the set $M_j(p)$ of *1-DIs* of cardinality of $w(M_j)=m$. Recall that here the set $M_j(p)$ represents the DeMorgan's clause $C_j(p)$ without specification of states (complemented or uncomplemented) of variables forming it. As it is easily seen, the procedure *Generate_M$_j$(p)* is of linear complexity in $n$. But it must be applied to each element of $S_{DM}(p)$ which would be of cardinality of $1 \leq w(S_{DM}(p)) \leq 2.5n$. Hence the worst-time complexity of application of this procedure is to be polynomial in $n$.

Note that we use the sets $M_1(p), M_2(p),\ldots, M_r(p)$ for generating $N(p)$-*vectors* to be used for producing PIs. Each $N(p)$-*vector* represents one certain PI generated by Nelson theorem. But in difference from an exact PI it shows only the positions of the literals that are to be appeared in this PI. For example, the $N(p)$-*vector 0101* for the function $f(x_1,x_2,x_3,x_4)$ indicates that there is a PI consisting of variables $x_2$ and $x_4$ but does not specify the states of these variables. As will be seen below, these states are clarified by bitwise ANDing each $N(p)$-vector with complement of the On-cube $P$.

The formula (3) states that to generate $N(p)$-*vectors,* it is sufficient to process the sets $M_1(p), M_2(p), \ldots, M_r(p)$ by the following iterative formula.

$$N(p)=N(p) \mid M_j(p), \quad \forall j \in \{1,2,3, \ldots ,w(S_{DM})\} \tag{6}$$

Where the initial state of $N(p)$ *is* $\{0\}^n$ and the bitwise OR operation ( $\mid$ ) on $N(p)$ and $M_j(p)$ may be performed as follows.

$$N(p) \mid M_j(p)=\{e_k \mid v_g : e_k \in N(p),\ v_g \in M_j(p)\}, \quad \text{where } k_{max}=w(N) \text{ and } g_{max}=w(M_j) \tag{7}$$

As it can be seen from the formula (7), an algorithm to be realizing the formula (6) is to be of polynomial complexity. But since it must be applied for $\forall M_j(p)$: $j \in \{1,2,\ldots, w(S_{DM})\}$ the application of that algorithm for $w(S_{DM})>>1$ is to be more complex than polynomial. So, if the cardinalities of the sets $M_j(p),M_2(p),\ldots,M_r(p)$ are $C_1, C_2,\ldots,C_r$, respectively, then according to the formula (7), the complexity of a algorithm realizing the formula (6) may be expressed as $C_1 \times C_2 \times \ldots \times C_r$, where $r=w(S_{DM})$. The calculations performed on a lot of logic functions show that this complexity can be approximated by the exponentional formula $C_{T}=O(2^{n-1})$. We are preventing this complexity by applying the formula (8).

$$\forall e_i,e_k \in N(p)(e_i \& e_k = e_i \rightarrow N(p)=N(p)\setminus e_k \quad and \quad e_i \& e_k = e_k \rightarrow N(p) = N(p)\setminus e_k ) \qquad (8)$$

The procedure *Minimize_N$_j$(p)* developed for realization of this formula removes redundant BSs from $N_j(p)$ as soon as they are generated. Based on the formula (7) and by using the procedure *Minimize_N(p)* we implement the formula (6) by the procedure *Extract_N$_j$(p)* given below.

***Extract_N$_j$(p)*** *(N(p), M$_j$(p), w(N), w(M$_j$))*

```
    k=1;  g=1;  q_j=0;  M_j(p)={0}^n
    For g=1 to w(M_j) Do
    {   For k=1 to w(N) Do
        {   e=e_k ¦ v_g;
            N(p)= N(p) ∪ e;
        }
       Minimize_N(p) (N(p), w(N))
    }
    Return (N(p),w(N)
```

Owing to procedure *Minimize_N(p)* the set *N(p)* grows so slowly that the procedure *Generate_N(p)* formed by binding the above explained procedures together remains in class of polynomial complexity in *n*.

***Generate_N(p)*** $(S_{DM}(p),w(S_{DM}))$

$N(p)=\{0\}^n$;

$r= w(S_{DM})$

*For j=1 to r Do*

{ *Read* $D_j \in S_{DM}(p)$

*Generate_*$M_j(p)$ $(D_{j,}\, n)$

*Extract_N(p)* $(N(p), M_j(p), w(N), w(M_j))$

}

*Return* $(N(p), w(N)$

***3.4. The Generation of Prime Implicants Covering the Given On-Cube***

In order to transform a vector $e \in N(p)$ into PI represented by this vector. It is sufficient to perform the operations: $C_L = \sim P \text{ ¦}(\sim e)$; $C_R = P \text{ ¦}(\sim e)$, where $(C_L, C_R)$ is the positional representation of the PI covering the On-cube *P*. Based on this transformation we can generate all PIs covering *P* by the following procedure that is of linear complexity in the number of PIs.

***Generate _*** $\boldsymbol{S_{PI}(p)}$ $(P, N(p), w(N))$

$S_{PI}(p)=\varnothing$;

*$P_L$= ~P; $P_R$ = P; i=1;*

*While i$\leq$w(N) Do*

*{ Read $e_i \in N(p)$;*

*e = ~$e_i$;*

*$C_{Lj} = P_L + e$;*

*$C_{Ri} = P_R + e$*

*$S_{PI}(p) = S_{PI}(p) \cup C_i$;*

*i=i+1*

*}*

*Return ($S_{PI}(p)$)*

### *3.5. The Main Procedure*

The main procedure *Generate_$S_{PI}$(p)* that to be generate all PIs for the given On-cube $P \in S_{ON}$ has been formed by sequencing the procedures given in this section.

***Generate_$S_{PI}$(p)*** *(P, $S_{OFF}$)*

*Generate_$S_{DM}$(p)(P, $S_{OFF}$, w($S_{OFF}$)   // Output: $S_{DM}$(p), w($S_{DM}$)*

*Generate_N(p)($S_{DM}(p)$, w($S_{DM}$))   // Output: N(p), w(N)*

*Generate _$S_{PI}$(p)(P, N(p), w(N))   // Output: $S_{PI}$(p)*

*END*

***Example.*** Find the complete set of PIs for the On-cube *P=11010* of the single-output function *f($x_1$,$x_2$,$x_3$,$x_4$,$x_5$)* represented by $S_{ON}$ = {00000, 00010, 00011, 01000, 01001,

01100, 01101, 01110, 10000, 10010, 11000, 11010, 11110} and $S_{OFF}$ = {00001, 00100, 00110, 01010, 01111, 10001, 10011, 10100, 10101, 10110, 10111, 11001, 11011, 11100, 11101, 11111}. Note that the reason for choosing the cube $P=11010 \in S_{ON}$ is that it causes appearance of all possible absorption relations that would take place between two cubes.

*Application of the procedure Generate_$S_{DM}(p)$*

In this part of the example the number of comparisons and the number of absorbed BSs (that scratched out) in the $i^{th}$ step are denoted by $C_i$ and $A_i$, respectively. The initial content of $S_{DM}(p)$ is $s_0= \{1\}^n$. The $j^{th}$ *DI* and the appropriate content of $S_{DM}(p)$ are formed by operations $D_j = P_R \oplus Z_j$ and $S_{DM}(p)=S_{DM}(p) \cup D_i$, respectively, where $Z_j \in S_{OFF}$.

$S_{DM}(p)=\{1\}^5=\{11111\}$

$D_1$= 11011; $S_{DM}$(p) = {~~11111~~, 11011}= {11011} // $C_1$=1, $A_1$=1

$D_2$= 11110; $S_{DM}$(p) = {11011, 11110} // $C_2$=1, $A_2$=0

$D_3$= 11100; $S_{DM}$(p) = {11011, ~~11110~~, 11100} // $C_3$=2, $A_3$=1

$D_4$= 10000; $S_{DM}$(p) = {~~11011~~, ~~11100~~, 10000} // $C_4$=2, $A_4$=2

$D_5$= 10101; $S_{DM}$(p) = {10000, ~~10101~~} // $C_5$=1, $A_5$=1

$D_6$= 01011; $S_{DM}$(p) = {10000, 01011} // $C_6$=1, $A_6$=0

$D_7$= 01001; $S_{DM}$(p) = {10000, ~~01011~~, 01001} // $C_7$=1 $A_7$=1

$D_8$= 01110; $S_{DM}$(p) = {10000, 01001, 01110} // $C_8$=2, $A_8$=0

$D_9$= 01111; $S_{DM}$(p) = {10000, 01001, 01110, ~~01111~~} // $C_9$=1, $A_9$=1

$D_{10}$= 01100; $S_{DM}$(p) = {10000, 01001, ~~01110~~, 01100} // $C_{10}$=3, $A_{10}$=1

$D_{11}$= 01101; $S_{DM}$(p) = {10000, 01001, 01100, ~~01101~~} // $C_{11}$=1, $A_{11}$=1

$D_{12}$= 00011; $S_{DM}$(p) = {10000, 01001, 01100, 00011} // $C_{12}$=3, $A_{12}$=0

$D_{13}$= 00001; $S_{DM}$(p) = {10000, ~~01001~~, 01100, ~~00111~~, 00001} // $C_{13}$=4, $A_{13}$=2

$D_{14}$= 00110; $S_{DM}$(p) = {10000, 01100, 00001, 00110} // $C_{14}$=3, $A_{14}$=0

$D_{15}$= 00111; $S_{DM}$(p) = {10000, 01100, 00001, 00110, ~~00111~~} // $C_{15}$=1, $A_{15}$=1

$D_{16}$= 00101; $S_{DM}$(p) = {10000, 01100, 00001, 00110, ~~00101~~} // $C_{16}$=2, $A_{16}$=1

Thus, $S_{DM}$(11010)={10000, 01100, 00001, 00110} and the average number of comparisons per Off-cube *is* $C=\sum_{i=1}^{16} C_i /k=29/16=1.81$.

*Application of the procedure Generate_N(p)*

$N_0$(p)={0}$^5$={00000}

$D_1$=10000 → $M_1$(p)={10000}

$N_1$(p)=$N_0$(p)¦$M_1$(p)=00000}¦{10000}={10000}

$D_2$=01100 → $M_2$(p)={01000, 00100}

$N_2$(p)=$N_1$(p)¦$M_2$(p)={10000}¦{01000, 00100}={11000, 10100}

$D_3$=00001 → $M_3$(p)={00001}

$N_3$(p)=$N_2$(p) ¦ $M_3$(p)={11000, 10100}¦{00001}={11001, 10101}

$D_4$=00110→ $M_4$(p)={00100, 00010}

$N_4$(p)=$N_3$(p)¦$M_4$(p)={11001,10101}¦{00100,00010}={~~11101~~,11011,10101,~~10111~~}

*Application of the procedure Generate_$S_{PI}$(p)*

There N(p)=$N_3$(p)={$e_1$, $e_2$}={11011, 10101} and ($P_L$,$P_R$)=(00101, 11010) due to P=11010. Therefore,

i=1

e = ~$e_1$ = ~11011=00100

$C_{1L}$=$P_{1L}$ ¦ e =00101 ¦ 00100=00101

$C_{1R}$=$P_{1R}$ ¦ e =11010 ¦ 00100=11110 } $C_1$=11x10

$S_{PI}$(p)={$C_1$}={11x10}

i=2

e = ~$e_2$ = ~10101=01010

$C_{2L}$=$P_{2L}$ ¦ e =00101 ¦ 01010=01111

$C_{2R}$=$P_{2R}$ ¦ e =11010 ¦ 01010=11010 } $C_2$= 1x0x0

Thus, $S_{PI}$(p) = {$C_1$,$C_2$} ={11x10, 1x0x0}

## 4. THE EXPERIMENTAL RESULTS

A lot of experiments were done to evaluate the runtime and quality of the results of the algorithm realizing the proposed method. The computer used was a PC with Intel® Core 2 Duo T7200 2.0 GHz and 1024 MB RAM. The quality of the results was measured by numbers of PIs (cubes) forming the minimized functions. Especially a set of 45 standard single-output MCNC benchmarks was solved by ESPRESSO-EXACT, ESPRESSO-SIGNATURE and by proposed method. The last two methods generated the same results and took approximately the same time. Therefore we are referring to both them as ESPRESSO. The results of solutions of 45 benchmarks are shown in Table

1. As seen from this table, for 16 benchmarks (group G1) the results generated by proposed method are significantly better than those obtained by ESPRESSO. But there are 2 benchmarks *m3* and *m4* (group G2) for which our method generated a little worse results than ESPRESSO. For all remaining 27 benchmarks (group G3) both methods obtained the same results. In general our method generated better, equivalent and worse result for 36%, 60% and 4% of the benchmarks, respectively. Our method has proved faster for 44 benchmarks by a factor of 2,7, on average, and a little slower for only one benchmark (*den*) for which it generated only 4 PIs instead of 14 ones generated by ESPRESSO. Notice that in the experiments we did not applied any ordering of the Onset and applied the simplest EPI identification rule that selects such a PI which covers more On-cubes than other ones. Namely, the quality of results generated by our method can be significantly improved by using a convenient ordering of the Onsets to be processed and a more sophisticated EPI identification rule given in [9,10,13,16,17,19].

## 5. CONCLUSION

In this study we propose a new approach that allows simultaneous computation of all PIs covering a given minterm of a given function to be minimized. This approach is based on the reduced Offset concept developed by *A.A. Malik, R.K. Brayton, A.R. Newton and A. Sangiovanni-Vincentelli in 1991*. The main difference between our approach and the prototype approaches is that we represent each reduced Off-cube by using only single n-bit string instead of 2n-bit one used in prototype method and in most of other ones. Such a representation of the reduced cubes allows us to speed up the reduced Offset generating process and to reduce memory amount required for this aim by the factors of 6 and 2, respectively. Because the proposed method generates all PIs covering the given On-cube simultaneously the direct-cover minimization algorithm based on this method is worked approximately 2.7 times faster, on average, than ESPRESSO. Our approach can also be applied to minimization of multiple-output functions by taking into consideration the well known relations existing between multiple-output PIs [15,18,19,21].

Table 1. The results of experiments on 45 standard single-output MCNC benchmarks

| | Benchmarks | n / $w(S_{ON})$ / $w(S_{OFF})$ | Number of result cubes | | Time elapsed (ms) | | $T_E$ / $T_{OM}$ |
|---|---|---|---|---|---|---|---|
| | | | Espresso | Our meth. | Espresso $T_E$ | Our Meth. $T_{OM}$ | |
| G1 | bca | 26/15/13 | 6 | 1 | 52,10 | 50,39 | 1,03 |
| | t10 | 10/134/189 | 49 | 45 | 50,67 | 22,43 | 2,26 |
| | br1 | 12/25/8 | 8 | 4 | 52,47 | 17,82 | 2,94 |
| | br11 | 12/28/5 | 8 | 3 | 50,45 | 31,88 | 1,58 |
| | br2 | 12/29/6 | 6 | 2 | 50,46 | 18,03 | 2,8 |
| | den | 18/18/2 | 14 | 4 | 57,13 | 76,93 | 0,74 |
| | exp | 8/18 /52 | 4 | 3 | 50,47 | 17,69 | 2,85 |
| | exp1 | 8/24/47 | 6 | 3 | 51,08 | 18,22 | 2,8 |
| | inc | 7/12/22 | 7 | 6 | 50,29 | 18,06 | 2,78 |
| | max4 | 9/37/8 | 36 | 6 | 68,91 | 20,03 | 3,44 |
| | min | 9/83/51 | 29 | 6 | 50,56 | 17,64 | 2,87 |
| | p82 | 5/11/13 | 5 | 4 | 51,47 | 17,83 | 2,89 |
| | pdc | 16/29/1891 | 11 | 4 | 52,36 | 17,78 | 2,94 |
| | prom2 | 9/142/145 | 8 | 7 | 51,57 | 17,63 | 2,93 |
| | spla | 16/67/2036 | 38 | 29 | 51,03 | 23,21 | 2,2 |
| | sqn | 7/48/48 | 12 | 8 | 51,12 | 17,81 | 2,87 |
| G2 | m3 | 8/98/30 | 14 | 16 | 54,97 | 18,43 | 2,98 |
| | m4 | 8/223/26 | 23 | 24 | 52,84 | 37,22 | 1,42 |
| | apex4 | 9/4/434 | 4 | 4 | 51,05 | 18,35 | 2,78 |
| | check | 4/4/9 | 1 | 1 | 47,22 | 18,13 | 2,6 |
| | check1 | 4/4/8 | 1 | 1 | 48,06 | 18,48 | 2,6 |
| | check2 | 4/4/6 | 1 | 1 | 49,97 | 18,14 | 2,75 |
| | check3 | 4/8/5 | 2 | 2 | 49,50 | 18,22 | 2,72 |
| | dist | 8/53/203 | 12 | 12 | 50,44 | 18,65 | 2,7 |
| | ex5 | 8/33/223 | 2 | 2 | 52,25 | 18,49 | 2,83 |
| | exps | 8/65/131 | 20 | 20 | 50,24 | 17,71 | 2,84 |
| | f51m | 8/128/128 | 23 | 23 | 53,30 | 18,07 | 2,95 |
| | linrom | 7/65/63 | 24 | 24 | 51,05 | 18,81 | 2,71 |
| | m | 6/27/5 | 4 | 4 | 49,97 | 17,74 | 2,82 |
| | m5 | 6/27/5 | 4 | 4 | 50,82 | 17,61 | 2,89 |
| | max1024 | 10/516/508 | 4 | 4 | 50,36 | 18,32 | 2,75 |
| **G3** | max128 | 7/29/99 | 8 | 8 | 52,38 | 18,09 | 2,9 |
| | max3 | 7/12/116 | 7 | 7 | 50,62 | 18,35 | 2,76 |
| | max512 | 9/358/254 | 10 | 10 | 51,02 | 17,66 | 2,89 |
| | mlp4 | 8/32/224 | 9 | 9 | 52,78 | 18,26 | 2,89 |
| | new2 | 6/3/4 | 2 | 2 | 52,84 | 18,72 | 2,82 |
| | poperom | 6/56/8 | 7 | 7 | 51,58 | 18,39 | 2,8 |
| | rd84 | 8/120/136 | 84 | 84 | 51,36 | 17,85 | 2,88 |
| | root | 8/15/241 | 4 | 4 | 53,15 | 17,91 | 2,97 |
| | sqr | 6/18/46 | 2 | 2 | 51,26 | 18,21 | 2,81 |
| | squar | 5/9/23 | 2 | 2 | 50,62 | 18,69 | 2,71 |
| | t3 | 12/27/121 | 6 | 6 | 51,82 | 17,83 | 2,91 |
| | wim | 4/9/1 | 4 | 4 | 51,27 | 18,12 | 2,83 |
| | z5xp1 | 7/25/103 | 3 | 3 | 50,16 | 17,91 | 2,8 |
| | e | 8/65/128 | 20 | 20 | 52,46 | 17,94 | 2,92 |

**APPENDIX A**. ***The procedure for generation of the set $S_{DM}(p)$***

***Reform_$S_{DM}$(p)***$(S_{DM}(p), D_j, w(S_{DM}))$

```
    Red_C =0
    While i ≤ w(S_DM) Do
    {     Read  s_i ∈ S_DM(p);
          A= s_i&D_j;
          If A=s_i Then
                Go to M
          Else If  A=D_j  Then
          {   s_i ={0}^n;
              Red_C= Red_C +1
          }
          i=i+1
    }
    S_DM(p) = S_DM(p) ∪ D_j;
    w(S_DM) = w(S_DM) +1
    If Red_C ≠0  Then
    {   Remove ∀ s ∈S_DM(p):s=0;
        w(S_DM) = w(S_DM) - Red_C
    }
    M:    Return (S_DM(p), w(S_DM))
```

If the last generated difference indicator $D_j$ is absorbed by $s_i \in S_{DM}(p)$ then the procedure *Reform_$S_{DM}$(p)* returns $S_{DM}(p)$ and its cardinality $w(S_{DM})$ without any change. If $D_j$ absorbs $s_i$ then $s_i$ is replaced by 0-BS ($\{0\}^n$), the content of the redundancy counter *Red_C* is increased by 1, and the *While–Do* loop is repeated with the next element of $S_{DM}(p)$. If there is no element in $S_{DM}(p)$ that can absorb $D_j$, then $D_j$ is included into $S_{DM}(p)$ and $w(S_{DM})$ is increased by 1. Then the content of *Red_C* is controlled. *Red_C*≠0 means that there are 0-BSs that are to be removed from $S_{DM}(p)$. Therefore in such a case the procedure removes all 0-BSs from $S_{DM}(p)$ and updates its cardinality.

**APPENDIX B.** ***The procedure for minimization of the set N(p)*** **on the formula (7).**

***Minimize_N(p)*** *(N(p),w(N))*

```
     i=1; k=2
R2:  While k ≤ w(N)  Do
     {   D = e_i&e_k;
         If  D=e_i then
               e_k=0
         Else If D=e_k  then
         {     e_i=0 ;
               Go to R1
         }
         Else
               k=k+1
```

*}*
*R1: i=i+1;*
*k=i+1*
*If i >( w(N) -1) then*
*Go to R3*
*Else If $e_i$ =0 then*
*Go to R1*
*Else*
*Go to R2*
*R3: N(p)= N(p)\{ $e_i$: $e_i$ =0} and update w(N)*
*Return (N(p), w(N))*

## APPENDIX C. *Some benchmarks for which our method gave much better results than ESPRESSO*

| **Bench-mark** | **n/ w($S_{ON}$)/ w($S_{OFF}$)** | **RESULTS**<br>**Results obtained by ESPRESSO**<br>**Results obtained by Our Algorithm** | | $w(S^E_{PI})/$ $w(S^O_{PI})$ |
|---|---|---|---|---|
| | | **Espresso** $S^E_{PI}$ | **Our Method** $S^O_{PI}$ | |
| Bca | 26/9/292 | 00000000000000000000000xxx<br>0000000000000000000000x000<br>1011000110x000000000000000<br>10x10010110000000000000000<br>x0000000000000000000000000<br>10110011000000000000000000 | x0xx00xxxxxxxxxxxxxxxxxxxx | 6/1 |
| br11 | 12/28/5 | 110xx10001x0<br>11x0x10001x0<br>110x1101110x<br>111111010xx0<br>111x010111x0<br>11x0010111x0<br>11xx01000100<br>111010111100 | xxxxxx0xxxx0<br>xxxx1x0xxxxx<br>xxx01xxxxxxx | 8/3 |
| den | 18/18/2 | 0110xx111111000001<br>01011x111111000001<br>0101x1111111000001<br>011x00111111000001<br>000000111001111111<br>000111111100000111<br>000001111001011001<br>000110111011001001<br>000011111010001110<br>000101111011001010<br>000010111010010010<br>010100111111000010<br>010001111110000010<br>010010111110000010 | xxxxxxxxxxx1xxx0xx<br>x0xxxxxxxxxxxxxx1x<br>xxxxx0xxxxxxxxx0xx<br>xxxx0xxxxxxxxxxx1x | 14/4 |
| min | 9/83/51 | 101100xxx 10110x1xx<br>11x0001xx 110xx011x<br>11x110x0x 1100xx10x<br>110xx01x1 1100xx1x0<br>1011110xx 1010100xx<br>1101x100x 11x000x11<br>1101x10x0 1101xx000<br>11010x111 x01101111<br>101010x00 001010111<br>000000011 010000001<br>001000100 110011xxx<br>11011x0xx 11100x01x<br>11100x0x1 1011xx111<br>1x1110110 11x001000<br>1x1110000 | x00xxx011<br>xx1x00100<br>xxxx10111<br>xxx10x111<br>x1xxxxx01<br>1xxxxxxxx | 29/6 |
| pdc | 16/29/1891 | 010000x001x00001<br>010000x00x100001<br>010000x001x00010<br>010000x00x100010<br>010000x001x00100<br>010000x00x100100<br>010000x010x00000<br>010000x01x000000<br>000000100x101010<br>000000100x000000<br>0000001010011010 | xxxxxx100x000000<br>x1xx00xxxxxxxxxx<br>xxxx0x100x101010<br>xxxx0x1x10011010 | 11/4 |
| max4 | 9/37/8 | x10000000 001101111<br>100001111 001110111<br>011010111 110000111<br>101101011 000101011<br>011001011 011110011<br>111100011 011101101<br>001100101 011000101<br>011111001 101111001<br>000111001 110101001<br>100001001 001010001<br>101000001 010111110<br>101011110 111001110<br>000011010 100101010<br>010110010 101010010<br>011100010 001011100<br>101001100 110110100<br>010100100 010111000<br>101011000 100110000 | xxx0x1xxx<br>xx01xxxx0<br>xxxxx00xx<br>xxxx0x0x1<br>1xx0xxx1x<br>0xxx11xxx | 36/6 |

## APPENDIX D. Generation of Multiple-Output Prime Implicants

In our opinion, among the existing methods generating the multiple-output PIs, the most suitably one, from the reduced Offset concept point of view, is the method proposed by Sharon R. PERKINS and Tom RHYNE [19]. In this study an On-minterm is denoted by TM (TRUE minterms) and TMs that make a given output function true are given its *Tags* (subscript), as $(3)_{2,3}$ which indicates that the minterm 3 (011) makes the functions $f_2$ and $f_3$ be true. Each tag is characterized by the number of functions appearing in it, which is called the weight of this tag. We denote the tag of the weight of m by *m-tag.*

TMs are ordered on their tags and TM having tag of smallest weight is selected first. The algorithm realizing this method is as follows [19].

***Procedure_EDSA*** ($S_{ON}$, $S_{DC}$}

1) SELECT a TM to be covered (the origin TM)
2) IDENTIFY the multiple–output PIs covering this TM
3) ELIMINATE PIs that are not needed to form a minimum cover
4) SELECT a PI to cover the origin TM (if possible)

UNTIL all TMs are covered.

Notice that our approach affects only the second step of this algorithm. So instead of generating PIs by the implicant expansion we compute them logically. But there is need to form the multiple-output function specified by TM to be handled. This is done by simply bitwise ANDing the output columns of the functions appearing in the tag of that TM. Then the generated multi-output function is processed as a single-output function but in respect only to TM at hand. As an example consider 3-input and 2-output function [18] given in the example below. In order to demonstrate how a multiple-output function is minimized we need to realize all of steps of the *Procedure_EDSA.* Therefore, in the example below we use some of concepts of minimization of multiple-output functions in spite of that they are out of scope of this study.

*Example.* Minimize the multiple-output function given in Table D1.

Table D1. Truth table of the multiple-output function to be minimized

| Inputs | | | Outputs (Tag) | | | Weight of the Tag |
|---|---|---|---|---|---|---|
| Q2 | Q1 | Q0 | Y2 | Y1 | Y0 | |
| 0 | 0 | 0 | 1 | 0 | 1 | 2 |
| 0 | 0 | 1 | 1 | 1 | 0 | 2 |
| 0 | 1 | 0 | 1 | 1 | 0 | 2 |
| 0 | 1 | 1 | 0 | 1 | 0 | 1 |
| 1 | 0 | 0 | 0 | 0 | 1 | 1 |
| 1 | 0 | 1 | 1 | 0 | 1 | 2 |
| 1 | 1 | 0 | 1 | 1 | 0 | 2 |
| 1 | 1 | 1 | 1 | 0 | 1 | 2 |

*The First Iteration*

As seen from Table D1, there are two TMs 011 and 100 with 1-tags (each of two tags of weights of 1). Since these tags (010, 001) are not intersecting (orthogonal), we can choose one of these TMs randomly. Let us firstly to choose TM $100_0$. This means that we have a sub function $Y_0(100)$ to be represented by $S_{ON0}(100)=\{100\}_0$ and $S_{OFF0}=\{001, 010, 011, 110\}_0$. By transforming $S_{OFF0}$ into $S_{MD}(100)_0$ by procedure *Generate_SDM(p)* We get the minimal set of difference indicators $S_{MD}(100)_0=\{101,010\}_0$ shown in Table D2.

Table D2. Generating DIs for the TM $100_0$

| $S_{OFF}(Y_0)$ | $S_{MD}(100)_0$ |
|---|---|
| 0 0 1 | 1 0 1 |
| 0 1 0 | ~~1 1 0~~ |
| 0 1 1 | ~~1 1 1~~ |
| 1 1 0 | 0 1 0 |

By transforming $S_{MD}(100)_0$ to $S_{PI}(100)_0$ by procedures *Extract_E(100)$_0$ and Generate_$S_{PI}$ (100)$_0$* we get the set of PIs covering TM $100_0$. This set is $S_{PI}(100)_0=\{10x_0,x00_0\}$. To make clear, whether there is an essential PI (EPI) or not, we obtain the subsets of TMs covered by PIs $10x_0$ and $x00_0$ as follows.

$M1(100)_0=S_{ON}(Y_0)\&10x=\{000, 100, 101, 111\}\&10x =\{100,101\}_0$
$M2(100)_0=S_{ON}(Y_0)\&x00=\{000, 100, 101, 111\}\&x00 =\{000,100\}_0$

The bit-string representations of these subsets are as follows.

$M1_{bs}(100)_0$= 0001100; $M2_{bs}(100)_0$= 1001000

Where $M1_{bs}(100)_0\&M2_{bs}(100)_0$=0001000. Since $0001000 \not\subset \{M1_{bs}(100)_0, M2_{bs}(100)_0\}$={0001100, 1001000}, there none of PIs is to be chosen as essential one.

We have another TM with 1-tag. It is $011_1$. But the tag of this TM does not intersect with the tag of TM $100_0$. Hence, processing of TM $100_0$ can be continued if there is not any TM with 1-tag. Otherwise we would stop the handling of TM $100_0$, continue with TM $011_1$ and go back to handling of TM $100_0$ if it is needed.

*Identification of the Essential PI by Heuristic Estimation of Affect of each PI to the Final Cover*

In order to choose one of PIs $10x_0$ and $x00_0$ as the essential one we have to examine the affect of these PIs to final cover. For this aim we use the neighbors of TM 100, which are obtained as follows.

$$N(100)_0 = (M1(100) \cup M2(100)) \setminus (M1(100) \cap M2(100))$$

Notice that in spite of non-simplicity of appearance of this expression it is implemented by single bitwise XOR operation as follows.

$$N_{bs}(100)_0= M1_{bs}(100)_0 \oplus M2_{bs}(100)_0=1000100$$

Here $N_{bs}(100)_0$=1000100 means that TM=100 has two neighbors consisting of 000 and 101. Both of these TMs belong to the sub function $Y_{2,0}$. Therefore we have to examine how PIs $10x_0$ and $x00_0$ affect this function. For this aim we have to examine the sub functions $Y_{2,0}$ (000) and $Y_{2,0}(101)$ that can not be appeared if we choose as the EPI $x00_0$ and $10x_0$, respectively. Therefore, we briefly explain the process of minimization both of these sub functions below. We form the $S_{OFF}$ of the sub functions $Y_{j,k,\ldots,l}$ as

$$S_{OFF}(Y_{j,k,..l}) =\{m \mid Y_j(m)\&Y_k(m)\&\ldots\&Y_l=0\}$$

where m is a minterm (TM) and $Y_i(m)$ is the value of the function $Y_i$ corresponding to the minterm m. Applying this rule to the first and third columns of the tag parts of Table D1 we obtain that $S_{OFF}(Y_{2,0})$ ={001, 010, 011, 100, 110}. Applying the procedure *Generate_SDM(p)* to this set for TMs $000_{2,0}$ and $101_{2,0}$ separately we get the results given in Tables D3 and D4.

Table D3. Generating DIs for the TM $000_{2,0}$

| $S_{OFF}(Y_{2,0})$ | $S_{DM}(000_{2,0})$ |
|---|---|
| 0 0 1 | 0 0 1 |
| 0 1 0 | 0 1 0 |
| 0 1 1 | ~~0 1 1~~ |
| 1 0 0 | 1 0 0 |
| 1 1 0 | ~~1 1 0~~ |

Table D4. Generating DIs for the TM $101_{2,0}$

| $S_{OFF}(Y_{2,0})$ | $S_{DM}(101_{2,0})$ |
|---|---|
| 0 0 1 | 1 0 0 |
| 0 1 0 | ~~1 1 1~~ |
| 0 1 1 | ~~1 1 0~~ |
| 1 0 0 | 0 0 1 |
| 1 1 0 | ~~0 1 1~~ |

By applying the procedures *Extract_E(p)* and *Generate_$S_{PI}$(p)* to the results given in Tables D3 (where $P=000_{2,0}$) and D4 (where $P=101_{2,0}$), we obtain that $Y_{2,0}(000) = 000$ and $Y_{2,0}(101) = 1x1$. There covering the TMs $000_0$ and $101_0$ causes to disappearing the sub functions $Y_{2,0}(000) = 000$ and $Y_{2,0}(101) = 1x1$, respectively. Since sub function $Y_{2,0}(101) = 1x1$ seems to be leading to a better result than the sub function $Y_{2,0}(000)=000$, we avoid of sub function $Y_{2,0}(000)=000$ by choosing $EPI_1=x00_0$. Meanwhile, we have obtained that the sub function $Y_{2,0}(101) = 1x1$ is to be part of good result. Therefore $EPI_2=1x1_{2,0}$ and $S_{PI} =\{x00_0, 1x1_{2,0}\}$. After covering Table D1 by these PIs it becomes as follows (Table D5).

Table D5. The state of Table D1 covered by PIs $x00_0$ and $1x1_{2,0}$

| Inputs | | | Outputs (Tag) | | | Weight of the Tag |
|---|---|---|---|---|---|---|
| $Q_2$ | $Q_1$ | $Q_0$ | $Y_2$ | $Y_1$ | $Y_0$ | |
| 0 | 0 | 0 | 1 | 0 | d | 1 |
| 0 | 0 | 1 | 1 | 1 | 0 | 2 |
| 0 | 1 | 0 | 1 | 1 | 0 | 2 |
| 0 | 1 | 1 | 0 | 1 | 0 | 1 |
| 1 | 0 | 0 | 0 | 0 | d | - |
| 1 | 0 | 1 | d | 0 | d | - |
| 1 | 1 | 0 | 1 | 1 | 0 | 2 |
| 1 | 1 | 1 | d | 0 | d | - |

*The Second iteration*

There are two TMs $000_2$ and $110_1$ with 1-tags in Table D5. Since the tags of these TMs are orthogonal (are not intersecting), we may select one of them randomly. Let us to chose TM $=000_2$. By applying the procedure *Generate_SDM(p)* to the sub function $Y_2(000)$ we get the set of difference indicators $S_{DM}(000_2)$ given in Table D6.

Table D6. Generating of DIs for the TM $000_2$

| $S_{OFF}(Y_2)$ | $S_{DM}(000_2)$ |
|---|---|
| 0 1 1 | 0 1 1 |
| 1 0 0 | 1 0 0 |

By applying the procedure *Extract_E(p)* and *Generate_$S_{PI}$(p)* to the result given in Table D6 we get $S_{PI}(000)_2$ ={00x, 0x0}. In order to make the existence of an essential PI clear, we obtain the subsets of TMs covered by PIs $00x_2$ and $0x0_2$ as follows.

$M1(000)_2=S_{ON}(Y_2)\&00x=\{000,001,010,110\}\&00x=\{000,001\}\rightarrow M1_{bs}(000)_2=11000000$
$M2(000)_2=S_{ON}(Y_2)\&0x0=\{000,001,010,110\}\&0x0=\{000,010\}\rightarrow M2_{bs}(000)_2=10100000$

But from $M1_{bs}(000)_2$ & $M1_{bs}(000)_2$ = 11000000 & 10100000 = 10000000 $\notin$ {11000000 & 10100000} follows that there none of PIs $00x_2$, $0x0_2$ can be identified as EPI. Therefore, we have to estimate these PIs indirectly-via the neighbors of TM $000_2$ for which they were generated.

Since $N_{bs}(000)_2= M1_{bs}(000)_2 \oplus M2_{bs}(000)_2=11000000 \oplus 10100000=01100000$, TM $000_2$ has two neighbors 001 and 010. As seen from Table D5, these TMs belong to the sub functions $Y_{2,1}(001)$ and $Y_{2,1}(010)$, respectively. Using Table D5 and procedure *Generate_SDM(p)*, the sets of $DI_S$ for the TMs $001_{2,1}$ and $010_{2,1}$ can be obtained as shown in Tables D7 and D8, respectively.

Table D7. Generating DIs for the TM $010_{2,1}$

| $S_{OFF}(Y_{2,1})$ | $S_{DM}(001_{2,1})$ |
|---|---|
| 0 0 0 | 0 0 1 |
| 0 1 1 | 0 1 0 |
| 1 0 0 | ~~1 0 1~~ |
| 1 0 1 | 1 0 0 |
| 1 1 1 | ~~1 1 0~~ |

Table D8. Generating DIs for the TM $001_{2,1}$

| $S_{OFF}(Y_{2,1})$ | $S_{DM}(010_{2,1})$ |
|---|---|
| 0 0 0 | 0 1 0 |
| 0 1 1 | 0 0 1 |
| 1 0 0 | ~~1 1 0~~ |
| 1 0 1 | ~~1 1 1~~ |
| 1 1 1 | ~~1 0 1~~ |

By applying the procedures *Extract_E(p)* and *Generate_$S_{PI}$(p)* to the results given in Tables D7 (where $P=001_{2,1}$) and D8 (where $P=010_{2,1}$) we obtain that $Y_{2,1}(001)$= 001; $Y_{2,1}(010)$= x10. Since the sub function $Y_{2,1}(010)$= x10 is to be leading to better result than the sub function $Y_{2,1}(001)$= 001 we must avoid of function $Y_{2,1}(001)$ by covering TM $000_2$ by PI $00x_2$. Thus $EPI_3=00x_2$, $EPI_4$= $x10_{2,1}$ and $S_{PI}$ ={$x00_0$, $1x1_{2,0}$, $00x_2$, $x10_{2,1}$}. After covering Table D5 by PIs $00x_2$ and $x10_{2,1}$ it becomes as follows (Table D9).

Table D9. The state of Table D5 covered by PIs $00x_2$ and $x10_{2,1}$

| Inputs | | | Outputs (Tag) | | | Weight of the Tag |
|---|---|---|---|---|---|---|
| $Q_2$ | $Q_1$ | $Q_0$ | $Y_2$ | $Y_1$ | $Y_0$ | |
| 0 | 0 | 0 | d | 0 | d | - |
| 0 | 0 | 1 | d | 1 | 0 | 1 |
| 0 | 1 | 0 | d | d | 0 | - |
| 0 | 1 | 1 | 0 | 1 | 0 | 1 |
| 1 | 0 | 0 | 0 | 0 | d | - |
| 1 | 0 | 1 | d | 0 | d | - |
| 1 | 1 | 0 | d | d | 0 | - |
| 1 | 1 | 1 | d | 0 | d | - |

Now we must handle one of TMs $001_1$ and $011_1$. But handling one of them by above-mentioned way shows that both of them are covered by the single PI $0x1_1$. Thus the minimal cover of multiple-output function represented by Table D1 is as

$S_{PI} = \{x00_0, 1x1_{2,0}, 00x_2, x10_{2,1}, 0x1_1\}$.

## APPENDIX D. Autobiographies of Authors

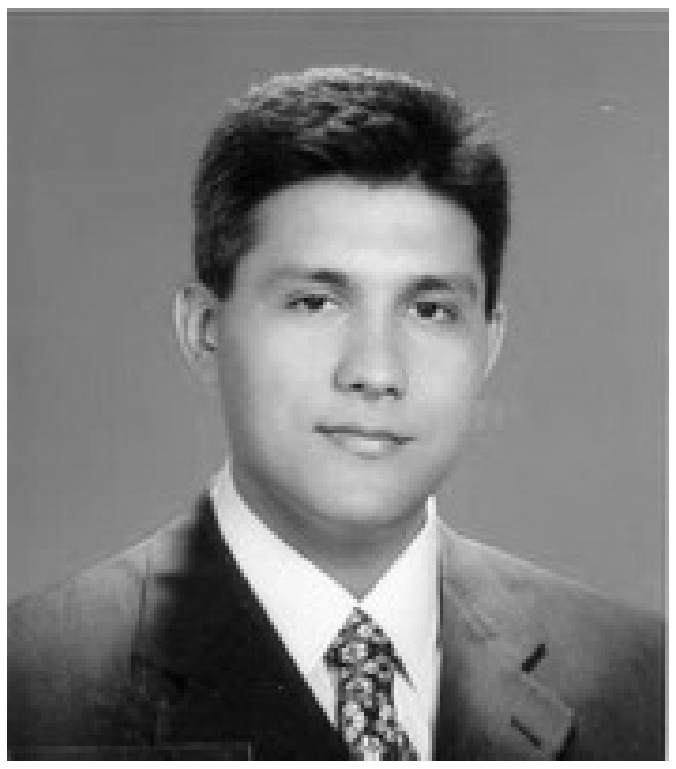

**Fatih Basciftci** was born in Konya, Turkey, in 1974. He received the Ph.D. degree in Electric and Electronic Engineering from Selcuk University, Konya in 2006. From 1998 to 2002 he was research assistant in Electronic and Computer Education Department of the same university. From 2002 to 2007 he was a lecturer of the same department of Selcuk University, Konya, Turkey. Since 2007 he has been an assist. professor of the same department. His research interest includes Switching Theory and Computer Architecture on which he has published over 20 papers.

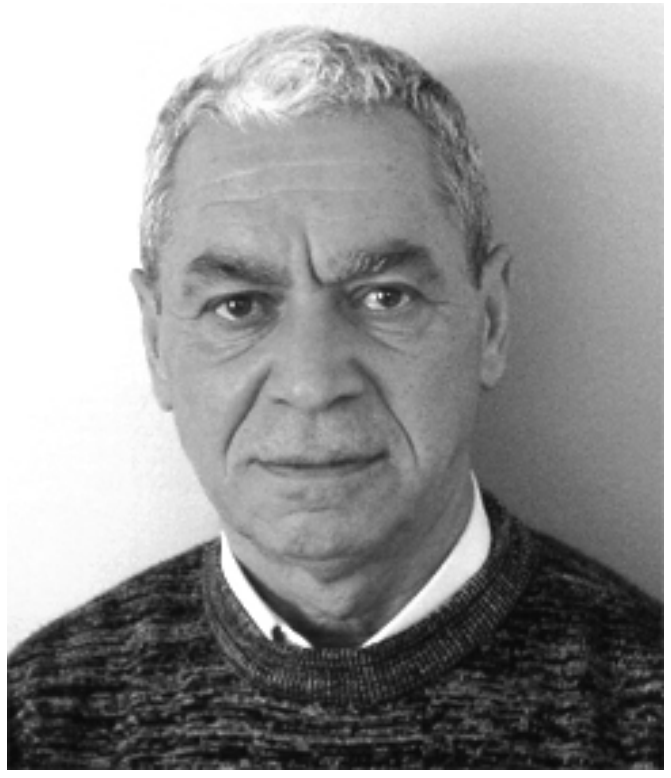

**Sirzat Kahramanlı** was born in Fizuli, Azerbaijan, in 1944. He received the Ph.D. degree in computer science from Azerbaijan Technical University, Baku in 1983. From 1968 to 1993 he worked in Computer System department of the same university. From 1993 to 2006 he was an associate professor of department of Computer Engineering of Selcuk University, Konya, Turkey. Since 2006 he has been a professor of the same department. His research interest includes Switching Theory and Computer Architecture on which he has published over 50 papers and several books (in Turkish).